\documentclass[prl,manuscript,aps,showpacs]{revtex4} 

\usepackage{graphicx}
\usepackage{dcolumn}
\usepackage{bm}
\usepackage{setspace}

\begin{document}

\title{Parkinson's Law Quantified: \\ Three Investigations on Bureaucratic Inefficiency}

\author{Peter Klimek$^{1}$, Rudolf Hanel$^{1}$, Stefan Thurner$^{1,2}$}
\email{thurner@univie.ac.at} 

\affiliation{
$^1$ Complex Systems Research Group; HNO; Medical University of Vienna; 
W\"ahringer G\"urtel 18-20; A-1090; Austria \\
$^2$ Santa Fe Institute; 1399 Hyde Park Road; Santa Fe; NM 87501; USA\\
} 


\begin{abstract}
We formulate three famous, descriptive essays of C.N. Parkinson on  
bureaucratic inefficiency  in a
quantifiable and  dynamical socio-physical framework.
In the first model we  show how the use of  recent opinion formation  
models for small groups
can be used to understand Parkinson's observation
that decision making bodies such as cabinets or boards become highly  
inefficient once
their size exceeds a critical 'Coefficient of Inefficiency', typically around 20.
A second observation of Parkinson -- which is sometimes referred to  
as Parkinson's Law --
is that the growth of bureaucratic or administrative bodies usually  
goes hand in hand with a
drastic decrease of its overall efficiency.
In our second model we view a bureaucratic body as a system of a flow  
of workers,
which enter, become promoted to various internal levels within the   
system over time,
and leave the system after having served for a certain time.  
Promotion usually is associated
with an increase of subordinates. Within the proposed model it  
becomes possible to
work out the phase diagram under which conditions bureaucratic growth  
can be confined.
In our last model we assign individual efficiency curves to workers  
throughout their life
in administration, and compute the optimum time to send them to old  
age pension, in order to
ensure a maximum of efficiency within the body -- in Parkinson's  
words we compute the 'Pension Point' . 

\end{abstract}


\maketitle

\section{I. Introduction}

C.N. Parkinson has studied socio-dynamical systems with special focus on bureaucratic and administrative bodies. In his famous book \cite{Park57} he comprises 10 essays which give detailed descriptions of microscopic mechanisms leading to macroscopic phenomena like exponential growth of administrations, impact on the size of executive councils on efficient decision-making, and the relationship between the age of retirement and a company's vital promotion scheme. These essays are not quantitative, however, the description of these microscopic mechanisms is so clear and brilliant, that they can be easily cast into dynamical models. We pick three of these essays which are especially well suited for this purpose.

In the first essay (chapter 4 in \cite{Park57}), Parkinson conjectured on historical evidence that there is a characteristic group size beyond which the ability of this group for efficient decision-making, i.e. the finding of consensus, is considerably diminished. The empirical findings based on government size, found in the 1950s \cite{Park57}, are still valid today to a large extend \cite{Klimek07}. By proposing an opinion formation model for small groups we provide an understanding for this hypothesis through the existence of finite size effects which lead to this characteristic size ('Coefficient of Inefficiency') of decision-making bodies.

In the second essay (chapter 1 in \cite{Park57}), we formalize the so-called 'Parkinson's Law'. We propose dynamical equations for his model of bureaucratic growth and solve them explicitly. As the main result we derive the phase diagram of this model for the macroscopic evolution of an administrative body. One phase corresponds to an exponentially growing body, the other to a shrinking one.

The third essay (chapter 10 in \cite{Park57}), studies the interrelation between the age of retirement and promotion schemes with the motif force to ensure the presence of career opportunities and prevent the lack thereof, a phenomenon also known as 'Prince Charles Syndrome'. There we show the existence of a unique maximum for the efficiency of a bureaucratic body at the optimum life-time of a worker in the system. Given a promotion scheme and the hierarchy in the system, the optimal age of retirement can be computed.

This paper is organized as follows: in Section II we review work on the relationship between the size of a decision-making body and its efficiency \cite{Klimek08}. In Section III we propose and solve the dynamical equations for a model of an administrative body which evolves according to Parkinson's Law. The question when an official within such a body should retire to ensure a maximum of overall efficiency of the administrative body is left for the final Section IV. We interpret our results and conclude in Section V.

\section{II. The Coefficient of Inefficiency}

Parkinson discovered what he called the life cycle of a cabinet. He studied the membership of Britain's highest executive council, the cabinet, spanning 700 years, from 1257 to 1955. He found that an initially small council steadily increased in size, until growing beyond a size of 20, shortly after which it was superseded by a new council with the same fate. The British cabinet went five times through this life cycle within the aforementioned timespan. He also lists the cabinet sizes of his time revealing that cabinets beyond a size of 21 are only found in communistic countries. He concludes that it is critical for cabinets to have memberships below this characteristic size (21) for efficient decision-making within this council (the 'Coefficient of Inefficiency'). His explanation for this phenomenon was that as group size increases the influence of individual members decreases (not only because there are simply more of them, but also because the group is more likely to dissociate into separate subgroups). The less influence a member bears, the more easily new members are admitted to the council which in turn decreases their influence further. In the following we ask if similar observations can be made with today's data on cabinets. We then propose a model for opinion formation in small groups whose dependence on the group-size nicely resembles the transition conjectured by Parkinson.

\begin{figure}[tbp]
 \begin{center}
 \includegraphics[height=58mm]{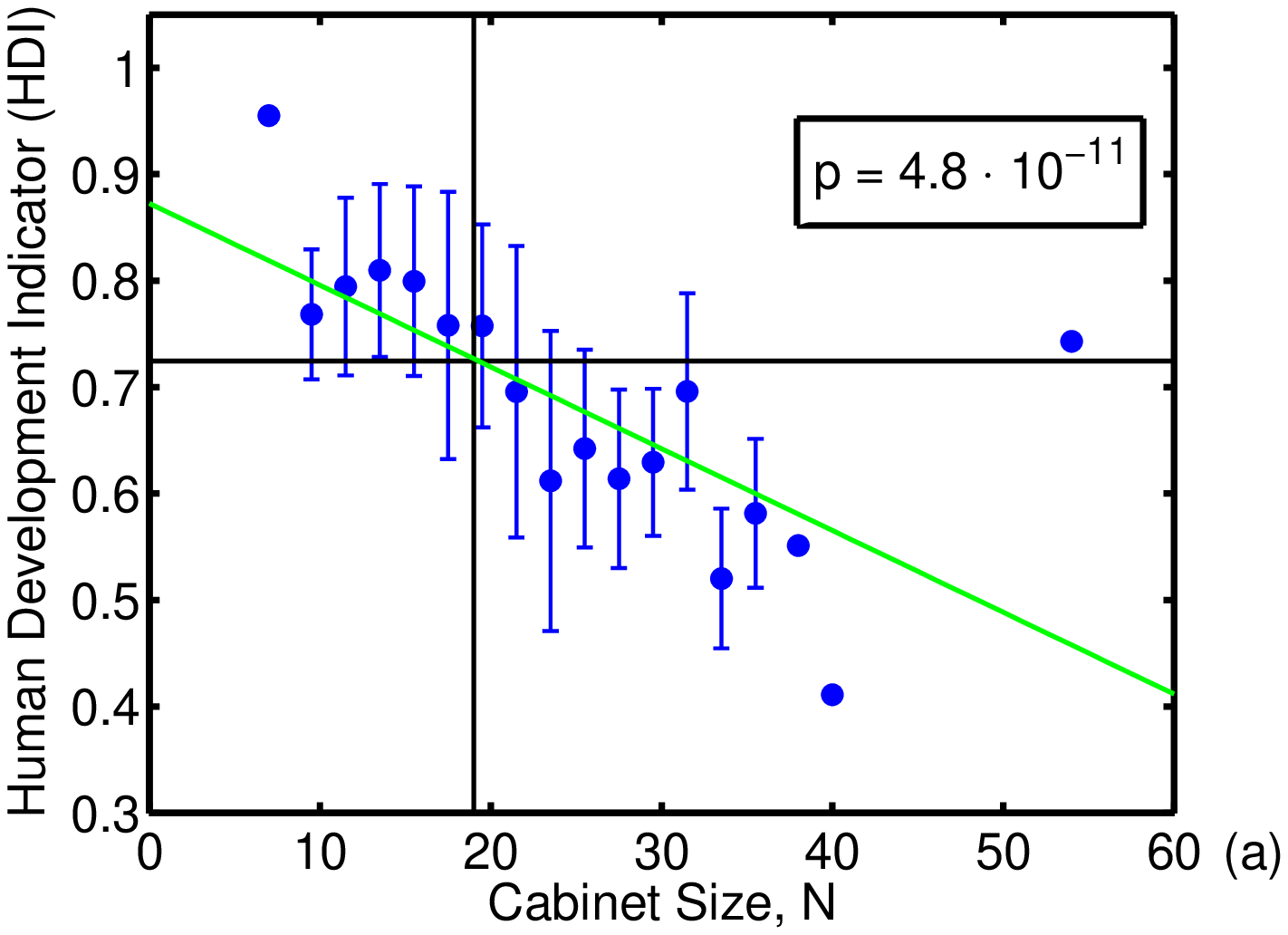}
 \includegraphics[height=58mm]{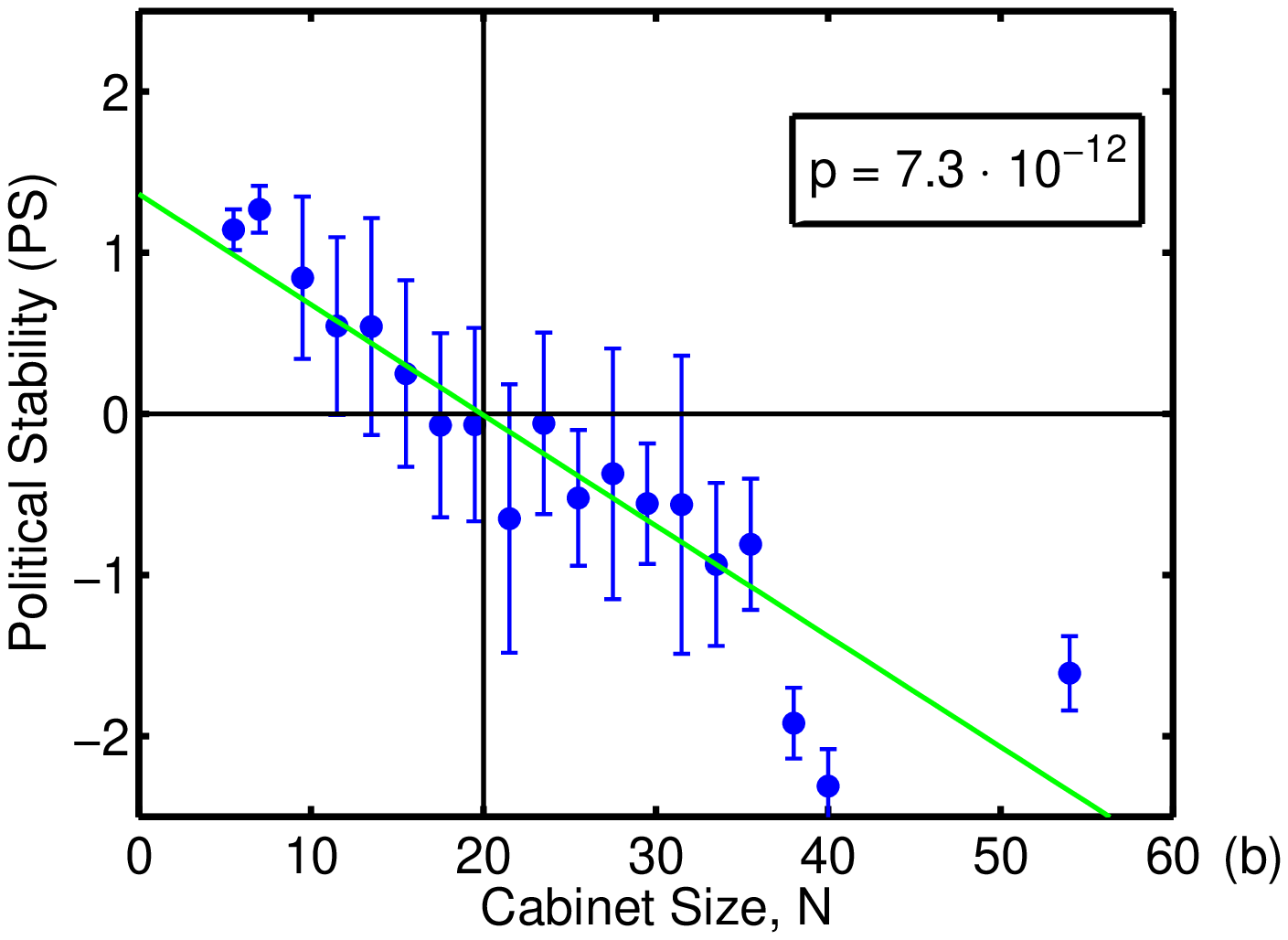}    
  \end{center}
 \caption{Cabinet size $N$ versus (a) the Human Development Indicator and (b) and Political Stability. The $p-$values indicate the high significance of negative correlation between them. We computed the average indicator value for each $N$ and binned them with intervals of two; errors were calculated by Gaussian error propagation. The green line is a linear least squares fit which intersects the median of the respective distributions at around 20.}
 \label{indplots}
\end{figure}

\subsection{Governmental Efficiency and Cabinet Size}

We compiled a database of cabinet sizes $N$, for 197 countries and self-governing territories with data drawn from the CIA \cite{chiefs}. The results reported in \cite{Klimek08} showed that the majority of cabinet sizes falls in the range between 13 and 20 \footnote{A curious observation was that there is no country with a cabinet consisting of eight members (though all other integers between 5 and 34 are found). Interestingly Parkinson reported the same observation in his investigations dating from 1957 \cite{Park57}.}. To compare cabinet size with its efficiency we made use of indicators provided by the United Nations Development Programme \cite{hdr} and the indicator Political Stability (PS) provided by the World Bank \cite{Kaufmann07}. We processed the data by computing the average of the indicator values for each $N$ and tested against the hypothesis that there exists no correlation between them, see Fig. \ref{indplots}. We find negative correlations between the cabinet size and the Human Development Indicator (HDI), see Fig. \ref{indplots}(a), and between cabinet size and a series of governance indicators from \cite{Kaufmann07} revealing that larger cabinets coincide with a more unstable political climate, Fig. \ref{indplots}(b). A linear fit to the data reveals that the value of $N$ separating above-average ranking countries from below-average ranking ones is found around 20 for \emph{each} indicator. For more details see \cite{Klimek08}.

\begin{figure}[tbp]
 \begin{center}
\includegraphics[height=50mm]{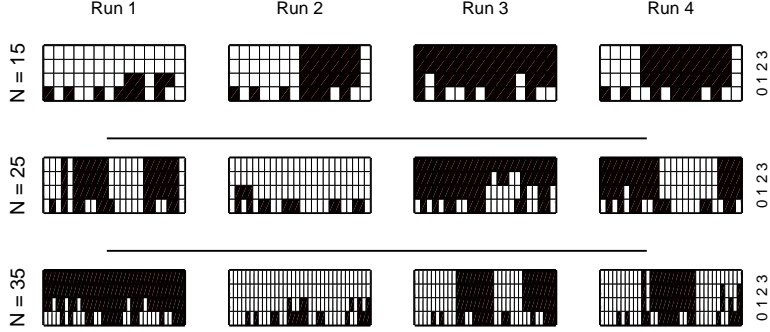}
 \end{center}
 \caption{Dynamics of the opinion formation process in groups of three different sizes ($N=15,25,35$). 
Each individual is a cell with binary opinion (black or white).
For each value of $N$ we show 4 independent update-runs, all starting with the same initial configurations 
and the same network. In each run we show the first four time steps 0,1,2,3, where time 0 corresponds to the initial
configuration. Line 1 is obtained by the iterative application of the opinion formation protocol on the 
initial configuration in a random sequence of updates.
The subsequent lines (2,3) are obtained analogously. The actual 
sequence is seen to play an important role for the final state at time 3. 
We show two different trajectories leading to consensus (everything white or black at time 3) 
or dissensus (mixed colors, i.e. opposing factions) for each size $N$.}
 \label{traj}
\end{figure}

\subsection{Opinion Formation in Small Groups}

Decision-making in small groups can be cast into a dynamical opinion formation model \cite{Liggett99}. Each member of the committee is represented as a node in a network and holds a binary opinion, say 0/1. Each node has a connectivity $k$ which is the number of undirected links to other nodes, representing a social influence (interactions, such as discussions) two agents exert upon each other \cite{Asch52, Kelman58}. Each node shares $k$ undirected links with other nodes. Thus for $N>k+1$ the graph is not fully connected and nodes appear which are not directly linked. We assume the total network to be connected, i.e. there exist no disjoint subnetworks. According to Parkinson groups such as cabinets are typically highly clustered. Therefore a sensitive choice for the fixed network architecture of our model is a small-world network, which has already been shown to be of paramount importance in modelling social interactions \cite{Watts99}. Each node is connected with its $k$ nearest neighbors, then with probability $e$ each link is randomly rewired. As a dynamical rule governing the interactions between connected agents we chose a majority rule \cite{BenNaim96,Axelrod97,Krapivsky03,Castellano05} with a predefined threshold $h\in \left(0.5,1\right]$, see \cite{Watts02}. When the update is carried out random sequentially, i.e. within one iteration each node is updated one after the other in a random order, this yields exactly the opinion formation model studied in \cite{Klimek07}. Here a node adopts the state of the majority of its $k$ neighbors only if this majority lies above $h$, otherwise the node keeps its previous internal state.

To anticipate the evolution of a system specified by the above characteristics, consider Fig. \ref{traj}, where four different runs for three choices of cabinet size $N$ are shown. We used parameter settings $k=8, h=0.6, e=0.1$. The system is initiated with a random configuration of internal states (lowest rows, labelled 0), each cell represents one agent in the network with a color corresponding to his binary opinion. The above rows are derived after updating each node once in a random sequential order. It becomes apparent that this model favors the forming of clusters, i.e. neighboring nodes tend to share the same opinion -- coalitions emerge. Note that for each of the four runs we start with the same initial conditions, thus the difference of the final states is solely due to the random sequential update. Further we see that the final states (rows number 3) can be characterized whether they consist of agents with the same opinion (consensus) or whether there are at least two factions of opposite opinions (dissensus).

\begin{figure}[tbp]
 \begin{center}
\includegraphics[height=60mm]{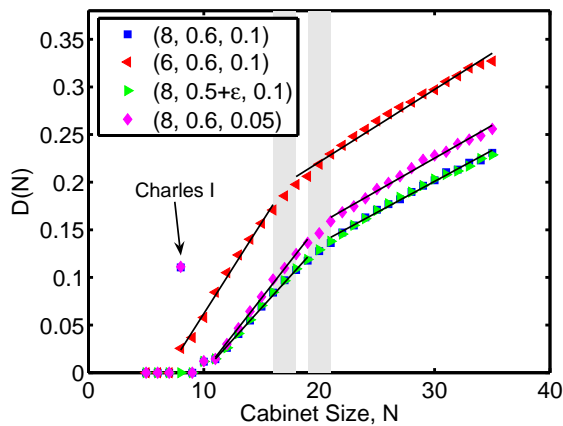}
 \end{center}
 \caption{$D\left(N\right)$ for four different settings of the model parameters $\left(k,h,e\right)$. Setting  $\left(8,0.6,0.1\right)$ $D\left(N\right)$ is shown by the blue squares. For $N<10$ we obtain always consensus here, except the $N=8$ case. We then find two regimes approximated by a linear fit and separated by a shaded area, where we conjecture the 'Coefficient of Inefficiency' to lie. Decreasing the connectivity of the group`s network, parameter setting $\left(6, 0.6,0.1\right)$, red left triangles, shifts the position of the shaded transition area and increases the tendency toward dissensus. Adjusting the threshold such that we recover the pure majority rule, $\left(8,0.5+\epsilon,0.1\right)$, green right triangles, has no impact on $D\left(N\right)$ due to our choice of $k$, except that we do not find the 'Charles I' scenario in this case. Lowering $e$ and therefore increasing the spatial correlations in the network hardens the finding of consensus too, as can be seen from the settings $\left(8,0.6,0.05\right)$ (magenta diamonds).}
 \label{DN}
\end{figure}

The question whether efficient decision-making in the group is possible or not translates into the question how likely the group evolves into final state of consensus or dissensus. We define dissensus $D\left(N\right)$ as the order parameter of the model: let $S_i$ be the initial population of nodes with internal state 0 and $S_f$ the final population in this state. We then have 
\begin{equation}
D\left( N \right) \equiv \left \langle \Theta \left(1- \frac{\max\left(S_f,N-S_f\right)}{N}\right) \right \rangle_{S_i}
\quad ,
\label{Dn}
\end{equation}
with $\Theta \left(x\right)$ being the Heaviside step function and $\langle \cdot \rangle_{S_i}$ the average over all possible initial conditions. $S_i$ is drawn with uniform probability from $\left(0,1,\dots N \right)$. Accordingly, the opinions are randomly assigned to the individual nodes. 
$D\left(N\right)$ gives the expectation value of a final state without consensus and measures the group's proneness to end up in dispute. We show $D\left(N\right)$ for four different parameter settings in Fig. \ref{DN}. For each curve we find three distinct regimes. In the region where the network is fully connected we always find consensus (except the notable $N=8$ case, to which we will return later). Then we find two regions of approximately linear growth. Between these, the shaded areas in the plot, the slope considerably decreases. We further see that for lower values of $k$, $D\left(N\right)$ increases (because of the emergence of disconnected clusters) and for less shortcuts, i.e. lower values of $e$ and therefore higher spatial correlations $D\left(N\right)$ also increases. Both mechanisms, lower $k$ and $e$, favor the emergence of clusters.

The decrease of the slope after $N$ has passed the sizes of the shaded areas can be naturally understood as a finite size effect present for small groups. This effect is constituted by the considerably increased likelihood to form coalitions in this model, is in one-to-one correspondence to the diminishing influence of the assembly's members. Since the gap between two values for $D\left(N\right)$ can be understood as the resistance to the cabinet's equivalent expansion, the dynamics proposed by Parkinson and the existence of a characteristic size for its functioning, the 'Coefficient of Inefficiency', are both reproduced by the behavior of $D\left(N\right)$ in Fig. \ref{DN}.

Parkinson has noted that there were no cabinets found with eight members. This is still the case today. Our findings show that there is good reason to avoid this number, at least for certain parameter values, see Fig. \ref{DN}. Historically there exists a famous exception: Eight was the number preferred by King Charles I for his Committee of State. And look what happened to him! \footnote{To learn more about Charles I, King of England, Ireland and Scotland -- and his decapitation -- visit http://www.royal.gov.uk/output/Page76.asp at the official website of the British monarchy.}.

\section{III. Parkinson's Law}

The famous sentence 'Work expands as to fill the resources available for its completion' was a guiding principle for Parkinson when he demonstrated that the growth of the administrative staff of the British navy stood in no relation to the work that actually had to be administered. For example, in the peacetime periods between 1935 and 1954 (i.e. excluding the even stronger growing staff during World War II) the staff of the Colonial office steadily increased by an average percentage of 6\% a year, from 372 to 1661, while the colonial territories in the same timespan shrunk dramatically. The same figure of 6\% turns out if one traces the growth of the dockyard staff of the British navy between 1914 and 1928. During this timespan the admiralty officials increased by around 80\%, while the ships in commission decreased from 62 to 20. Parkinson suggested to explain this growth by the 'principle' that an official tries to maximize subordinates, not rivals. This is nowadays soetimes referred to as Parkinson's Law. That is, if an official gets promoted (happening with probability $p$) this is in real terms carried out by allocating $r$ subordinates to him. 

We now present a model to understand Parkinson's intuition for administrative growth. We derive the growth rate $\lambda$ explicitly, depending on the model parameters $r$ and $p$. Further we introduce a drop-out rate $\gamma$, the probability for an official to quit his job, as well as the age of retirement $\tau_R$, which gives the typical administrative life-span. Consider $\gamma$ and $\tau_R$ to be fixed and measurable. The question to determine $\tau_R$ is left to the next section.

Let us denote the number of promoted staff that has served for $\tau$ years at time $t$, $n_{\tau}^+\left(t\right)$ and the number of subordinate staff $n_{\tau}^-\left(t\right)$, analogously. Let the age of retirement be $\tau_R$, the time between entry and end of service. We then have for the number of total promoted/subordinate staff $N^{\pm}\left(t\right)= \sum_{\tau=0}^{\tau_R} n_{\tau}^{\pm}\left(t\right)$. Further, let $p_{\tau}$ be the probability to get promoted after $\tau$ years of service, $r$ the average number of subordinates one gets appointed if promoted, and $\gamma$ the drop-out rate. We then arrive at the update equations
\begin{eqnarray}
n_{\tau}^+\left(t+1\right) & = & n_{\tau}^+\left(t\right) \left(1-\gamma\right) + p_{\tau} n_{\tau}^-\left(t\right) 
\quad, \label{D1} \\
n_{\tau}^-\left(t+1\right) & = & n_{\tau}^-\left(t\right) \left(1-\gamma-p_{\tau}\right)
\quad, \label{D2} \\
n_{0}^-\left(t+1\right) & = & r \sum_{\tau=0}^{\tau_R-1} p_{\tau} n_{\tau}^-\left(t\right)
\quad. \label{D3}
\end{eqnarray}
Assume that no official enters at a promoted level, $n_{0}^+\left(t\right)=0$. By introducing $p$ as the average probability to get promoted, $p=\frac{\sum_{\tau=0}^{\tau_R-1} p_{\tau} n_{\tau}^-\left(t\right)}{\sum_{\tau=0}^{\tau_R-1} n_{\tau}^-\left(t\right)}$, Eq. \ref{D3} takes the form
\begin{equation}
n_{0}^-\left(t+1\right) = r p \left(N^-\left(t\right)-n_{\tau_R}^-\left(t\right)\right)
\quad. \label{D3new}
\end{equation}
Combining this with Eq. \ref{D2} in the definition of $N^-\left(t\right)$ we obtain the update equation for the total number of subordinate staff,
\begin{eqnarray}
N^-\left(t+1\right) & =& \left(1-\gamma+\left(r-1\right)p\right)\left[N^-\left(t\right)-n_{\tau_R}^-\left(t\right)\right] \nonumber \\ 
 & = & \left(1-\gamma+\left(r-1\right)p\right) \left(1-q_-\right) N^-\left( t \right)
\quad, \label{N_min}
\end{eqnarray}
where we define $q_{-}=\frac{n_{\tau_R}^{\pm}\left(t\right)}{N^{-}\left(t\right)}$ and assume that the growth of the system is stationary. Using Eq. \ref{D1} in the definition for $N^{+}\left(t\right)= \sum_{\tau=0}^{\tau_R} n_{\tau}^{+}\left(t\right)$ we get the update equation for the total number of promoted staff as,
\begin{eqnarray}
N^+\left(t+1\right) & =& \left(1-\gamma\right) \left[N^+\left(t\right)-n_{\tau_R}^+\left(t\right)\right]+p \left[N^-\left(t\right)-n_{\tau_R}^-\left(t\right)\right] \nonumber \\
 &=& \left(1-\gamma\right) \left(1-q_+\right) N^+\left(t\right) + p \left(1-q_-\right) N^-\left(t\right)
\quad, \label{N_plu}
\end{eqnarray}
where $q_{+}=\frac{n_{\tau_R}^{\pm}\left(t\right)}{N^{+}\left(t\right)}$. We define $a\equiv\left(1-\gamma+\left(r-1\right)p\right) \left(1-q_-\right)$, $b\equiv p\left(1-q_-\right)$ and $c\equiv\left(1-\gamma\right)\left(1-q_+\right)$ to rewrite Eqs. \ref{N_min}, \ref{N_plu} in matrix form
\begin{equation}
 \left(\begin{array}{c} N^-\left(t+1\right) \\ N^+\left(t+1\right) \end{array} \right) = \left(\begin{array}{cc} a & 0 \\ b & c \end{array} \right) \left(\begin{array}{c} N^-\left(t\right) \\ N^+\left(t\right) \end{array} \right)
\quad. \label{N_MF}
\end{equation}
To obtain $q_-$ let us rewrite Eq. \ref{N_min} as $N^-\left(t+1\right)=a N^-\left(t\right)$. From Eq. \ref{D2} we find that $n_{\tau_R}^-\left(t\right) = n_{0}^-\left(t-\tau_R\right) \left(1-\gamma-p_{\tau}\right)^{\tau_R}$. Here we can plug in Eq. \ref{D3}, iterate $N^-\left(t-\tau_R-1\right)$ forward for $\tau_R+1$ times and divide by an overall $N^-\left(t\right)$ to obtain a self-consistent equation for $q_-$,
\begin{equation}
q_- \left(1-q_-\right)^{\tau_R}= \frac{r p \left(1-\gamma -p\right)^{\tau_R}}{\left[1-\gamma+\left(r-1\right)p \right]^{\tau_R+1}}
\quad. \label{qmin}
\end{equation} 
For $q_+$ we start with Eq. \ref{D1} and iterate it $\tau_R$ times backward. Making use of Eqs. \ref{D2} and \ref{D3} and using $n_{0}^+\left(t\right)=0$ we get
\begin{equation}
n^+_{\tau_R} \left( t\right)=p \sum_{t=0}^{\tau_R-1} n^-_{\tau} \left(t-\tau_R+\tau\right) \left(1-\gamma\right)^{\tau_R-\tau+1}
\quad. \label{qpl01}
\end{equation}
Applying Eq. \ref{D2} for $\tau$ times we obtain a sum which can be written as an incomplete geometric series leading to
\begin{equation}
n^+_{\tau_R} \left( t\right)=p \left( 1-\gamma\right)^{\tau_R+1} n^-_{0} \left( t-\tau_R\right)\frac{\left(\frac{1-\gamma-p}{1-\gamma}\right)^{\tau_R}-1}{\frac{1-\gamma-p}{1-\gamma}-1}
\quad. \label{qpl02} 
\end{equation}
Combining Eq. \ref{D3} and Eq. \ref{N_min} we get the helpful relation $
n_0^-\left(t-\tau_R\right) = r p \left(1-q_-\right) a^{-\tau_R-1} N^-\left( t\right)$ Plug this in Eq. \ref{qpl02} and divide both sides by $N^+\left( t\right)$ to finally obtain 
\begin{equation}
q_+=\frac{1}{\mu} \frac{rp\left(1-\gamma\right)^{\tau_R}\left[1-\left(1-\frac{p}{1-\gamma}\right)^{\tau_R}\right]}{\left[1-\gamma+\left(r-1\right)p\right]^{\tau_R+1} \left(1-q_-\right)^{\tau_R}} 
\quad, \label{qpl}
\end{equation}
where $\mu$ is the ratio of promoted to subordinate staff $\mu=\frac{N^+\left(t\right)}{N^-\left(t\right)}$. Let $\lambda$ be the growth rate of the entire staff $\left[N^+\left(t\right)+N^-\left(t\right)\right] \propto \mathrm e^{\lambda t}$.  Parkinson's Law then takes the form of the eigenvalue problem of a triangular matrix
\begin{equation}
\lambda \left(\begin{array}{c} 1 \\ \mu \end{array} \right) = \left(\begin{array}{cc} a & 0 \\ b & c \end{array} \right) \left(\begin{array}{c} 1 \\ \mu \end{array} \right)
\quad. \label{PL}
\end{equation}
We find the eigenvalues immediately in the diagonal. This system allows one meaningful solution for $\lambda$ and $\mu$, namely $\lambda=a$ and $\mu=\frac{b}{a-c}$. For the other solution, $\lambda=c$, $\mu$ diverges. 

\begin{figure}[tbp]
\begin{center}
\includegraphics[height=60mm]{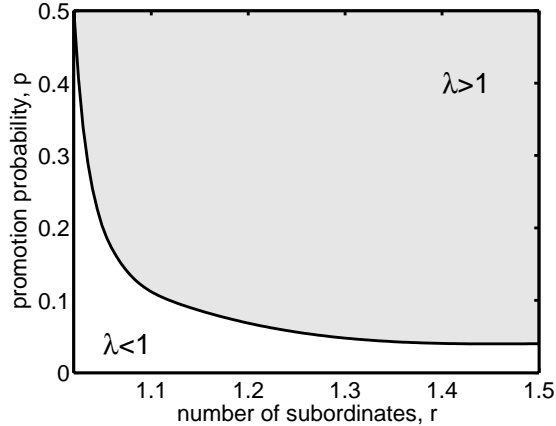}
 \end{center}
 \caption{Phase diagram for an administrative body evolving according to Parkinson's Law, Eq. \ref{PL} with  $\gamma=0.01$. The solid line (given by $\lambda=1$) separates a regime of a exponential body (white area) from shrinking growth (grey area).}
\label{lambda}
\end{figure}

We are finally in the position to answer the century, if not millennium old question how a system of bureaucrats can be organized such that it does \emph{not} grow exponentially. The phase diagram in Fig. \ref{lambda} shows the two phases, one of exponential growth, one of shrinking, for administrative bodies evolving according to Parkinson's Law. The line separating these two phases (given by $\lambda=1$) shows how the parameters $r$ (number of subordinates) and $p$ (average promotion probability) should be adjusted to keep the size of the bureaucracy constant (drop-out rate was set to $\gamma=0.01$).

\section{IV. Age of Retirement}

To fully understand Parkinsonian growth of administrations we now determine the age of retirement $\tau_R$. Parkinson suggests that the optimal value for $\tau_R$ has to be found on the basis of an administrative body's promotion scheme. The age of retirement thus does not depend on the actual age of the official whose retirement we are considering, but on his possible successors which should be given the possibility to advance in their careers. If this is not ensured a mechanism nowadays called 'Prince Charles Syndrome' sets in, i.e. the official gets frustrated by a lack of promotion opportunities, and individual efficiency decreases.

\begin{table}[tbp]
\caption{Stages of efficiency of an official. He enters at stage 1. After the respective number of years in service (number in brackets next to Parkinson's Coding) he advances to the next stage. If he does not reach a promoted position by reaching stage 6, he will enter the career described by the right column. We also show the values of the efficiency for official $i$, $\epsilon_i$, we assign to each stage.\\}
\begin{tabular}{rlclc}
\vspace{1mm}
\parbox{9mm}{Stage\\ \quad} & \parbox{41mm}{ Parkinson's Coding \\ (promotion at stage 5)} & \parbox{30mm}{Efficiency $\epsilon_i$ \\(promotion at 5)} & \parbox{36.5mm}{Parkinson's Coding\\ (no promotion)} &\parbox{27.5mm}{Efficiency $\epsilon_i$ \\(no promotion)} \\

\hline

1 & Age of Qualification (0) & $0.1$  & -- & $0.1$ \\
2 & Age of Discretion (3) &  $0.2$& -- & $0.2$ \\
3 & Age of Promotion (10) & $0.3$&-- & $0.3$ \\
4 & Age of Responsibility (15) &$0.4$ &--&$0.4$ \\
5 & Age of Authority (18)&$0.5$ &--&$0.5$ \\
6 & Age of Achievement (25) & $0.6$ &Age of Frustration (25) &$0.4$\\
7 & Age of Distinction (34) &$0.7$&Age of Jealousy (34) &$0.3$ \\
8 & Age of Dignity (40)&$0.8$ &Age of Resignation (38) &$0.2$ \\
9 & Age of Wisdom (43) &$0.9$ &Age of Oblivion (43) &$0.1$ \\
10 & Age of Obstruction (50) &$0.0$& -- &$0.1$ 
\end{tabular}
\label{eff}
\end{table}

We assume the hierarchical structure of the administrative body to be built according to the motif force that 'Officials try to maximize subordinates, not rivals', i.e. each official seeks to achieve a number $r$ of subordinates. One can picture this structure as a pyramid consisting of horizontal layers of officials holding an equally high office. On the hierarchical level above there are then $1/r$ times less, on the level below $r$ times more officials. Let us label the levels in the pyramid by an index $l$, then there are $r^l$ officials at level $l$. If we fix the number of levels to $L$ we have $N=\sum_{l=0}^{L} r^l$ and $l \in \{0,\dots,L\}$. We assign to each official $i$ an efficiency $\epsilon_i$ which depends on the years he has served and his current position. We use the career stages given as in Tab. \ref{eff}, and assign an $\epsilon_i$ to each stage, increasing linearly. These career stages are used as a tribute as in \cite{Park57}, but one could equally well work with other efficiency curves. The only important feature is the existence of a bifurcation point reflecting the age where the Prince Charles Syndrome sets in, here this is stage 6. At stage 1 we have $\epsilon_i=0.1$. For each subsequent stage (left column) it increases by $0.1$. If after reaching stage 6 the official $i$ reached a promoted position his efficiency proliferates. If at stage 6 we still find him in a subordinate position, his efficiency will decline by $0.1$ in each of the following stages (right column). We introduce the parameter $L'$ which is the lowest level which is counted as promoted. The subordinate staff is then present on the levels $L'<l\leq L$. The number of subordinate levels is called $\Delta L = L-L'$.

\begin{figure}[tbp]
\begin{center}
\includegraphics[height=60mm]{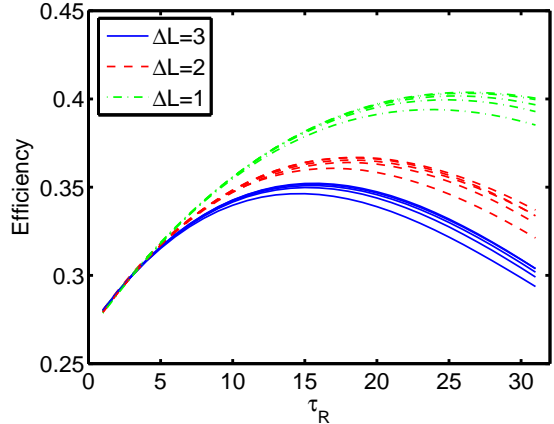}
 \end{center}
 \caption{Efficiency of a Parkinsonian administrative body versus age of retirement $\tau_R$. For three values of $\Delta L$ ($\Delta_L=1$, green slash-dotted, $\Delta_L=2$, red slashed, $\Delta_L=3$, blue solid line) we show results for systems of sizes $L=3,4,5,6,7$ (curves from bottom to top). With increasing system size the efficiency becomes size-independent. There is a unique maximum which position depends on $\Delta L$ for $\tau_R$.}
\label{PP}
\end{figure}

We numerically study this system in the following way. It is initialized with randomly distributed ages (and therefore efficiencies) and $r=2$. One time step in the simulation corresponds to one year. At each iteration each official ages by one year and his efficiency is updated according to Tab. \ref{eff}. If he reaches the age $\tau_R$ he retires and his position becomes vacant. One of his two subordinates is chosen randomly to replace him, leaving a vacant position one level below, triggering a cascade of promotions. When the lowest hierarchical level $L$ is reached the vacant position is taken by a new official who enters the system at stage 1. This process is iterated until the average efficiency in the system -- $\langle \epsilon_i \rangle_N$, the mean of $\epsilon_i$ over all officials $i=1,\dots,N$ -- approaches a stationary value. Simulation results are shown in Fig. \ref{PP}. We vary $\tau_R$ and compute the average efficiency $\langle \epsilon_i \rangle_N$ for systems of sizes $L=3,\dots,7$ and $\Delta L=$1 (green slash-dotted lines), 2 (red slashed lines), 3 (blue solid lines). For increasing $L$ the system approaches a stationary state where the efficiency only depends on $\Delta L$. Accordingly we find three bundles according to $\Delta L=1,2,3$ with curves for different values of $L$ discernible. Curves of same color and style correspond to increasing $L$ from bottom to top.
For each bundle we find a unique maximum indicating the optimal choice for $\tau_R$. For values of $\tau_R$ below this maximum officials retire too early to reach higher efficiencies, above this point the Prince Charles Syndrome becomes a danger. We further see that the flatter the hierarchy, i.e. higher $\Delta L$, the lower the optimal age of retirement.
Recall that we defined the ratio between promoted and subordinate staff as $\mu=\frac{N^+\left(t\right)}{N^-\left(t\right)}=\frac{\sum_{l=0}^{L-\Delta L-1} r^l}{\sum_{l=L-\Delta L}^{L} r^l}$.  By re-labelling the summation index we can write this as $\left(\sum_{l=0}^{\Delta L} r^l\right)^{-1} \sum_{l=1}^{L-\Delta L} \left( \frac{1}{r}\right) ^l$ which for large $L$ and $r>1$ converges to
\begin{equation}
 \mu=\left(\sum_{l=0}^{\Delta L} r^l\right)^{-1}
\quad. \label{mu_DL}
\end{equation}
For large system sizes the ratio $\mu$ indeed approaches a stationary value. So $\Delta L$ determines $\mu$ and the best choice for $\tau_R$, which in turn can be used to compute the growth rate $\lambda$ from the previous model.

\section{V. Conclusions}

We quantified three famous, descriptive essays of C.N. Parkinson on  
bureaucratic inefficiency  in a dynamical socio-physical framework.
In the first model we showed how the use of  recent opinion formation  
models for small groups
could be used to understand Parkinson's observation
-- which we showed is still valid in modern data --
that decision making bodies such as cabinets or boards become highly  
inefficient once
their size exceeds a critical 'Coefficient of Inefficiency'.
We showed how this characteristic size arises due to finite-size effects
in the process of forming of coalitions in small systems.
Parkinson's Law states that the growth of bureaucratic or administrative bodies usually  
goes hand in hand with a
drastic decrease of its overall efficiency.
In our second model we pictured a bureaucratic body as a system of a flow  
of workers through an administrative body. Officials enter, become promoted to various internal levels within the   
system over time,
and leave the system after having served for a certain time, they retire.  
Within the proposed model we showed how to
compute the phase diagram under which conditions bureaucratic growth  
can be confined. We thereby link Parkinson's microscopic interaction rules
to macroscopic properties observable in administration and bureaucracy.
It is possible to adjust the 'microscopic' model parameters -- which are altogether observable and measurable --,
the individual promotion probability and number of subordinates,
to control the growth rate of the bureaucratic body.
In our last model we assign individual efficiency curves to workers  
throughout their life
in administration, and compute the optimum time to send them into 
retirement, in order to
ensure a maximum of efficiency within the body, Parkinson's  
'Pension Point'.
We implement Parkinson's observation that an individual's efficiency 
declines if he resides in a subordinate position for too long,
i.e. the individual efficiency curves over time are characterized by the existence of bifurcation points.
This effect necessitates the need of promotion opportunities,
which can be ensured by two mechanisms here.
On the one hand one can lower the age of retirement allowing young
workers to advance in their career faster, or on the other hand one can
increase the levels of internal hierarchy.

\end{document}